\documentclass{elsart}

\usepackage{graphicx,amssymb}

\newcommand\myFootnote[1]{%
  \textsuperscript{\scriptsize\refstepcounter{footnote}\thefootnote}%
  \footnotesize #1\normalsize%
}

\begin{document}

\begin{frontmatter}



\title{Neutron Flux at the Gran Sasso Underground Laboratory Revisited}


\author{H. Wulandari\corauthref{cor1}},
\ead{Hesti.Wulandari@ph.tum.de} \corauth[cor1]{Corresponding
author. Tel: +49-89-28914416, Fax: +49-89-28912680.}
\author{J. Jochum},
\author{W. Rau},
\author{F. von Feilitzsch}

\address{Physikdept. E-15, Technische Universit\"{a}t M\"{u}nchen, James-Franck-Str. \\85748 Garching, Germany}

\begin{abstract}
The neutron flux induced by radioactivity at the Gran Sasso
underground laboratory is revisited. We have performed calculations and Monte Carlo simulations;
the results offer an independent check to the available experimental data reported by
different authors, which vary rather widely. This study gives
detailed information on the expected spectrum and on the variability of the
neutron flux due to possible variations of the water content of
the environment.
\end{abstract}

\end{frontmatter}

\section{Introduction}
One of the important parameters characterizing an underground
environment is the neutron flux. Recently, the role of neutrons as
a source of background has received more attention. Estimates
of the neutron flux from surrounding materials are needed for rare
event searches in underground sites. Highly sensitive underground
neutrino experiments require precise knowledge of neutron-induced
background in the detector. In direct searches for WIMP dark
matter neutrons are a particularly important background, because
they interact with the detector in the same way WIMPs do.

The dominant sources of neutrons at large depth underground, where
the cosmic rays are reduced significantly, are the ($\alpha$,n)
reactions on light elements (e.g. Li, F, Na, etc.) and the
spontaneous fission, mainly of $^{238}$U. In general, at a depth
of 3000$\,-\,4000$ m w.e., where many rare event experiments are
located, the flux of neutrons from activity of the experimental
environment is two to three orders of magnitudes higher than the
flux of neutrons from cosmic ray muons.

Due to its very low intensity, the neutron flux underground is not
easy to measure. At the Gran Sasso underground laboratory the
neutron flux has been measured by several groups, employing
different methods of detection
\cite{aleksan,arneodo,belli,bellotti,cribier,rindi}. To get the neutron flux from the measurements,
Monte Carlo simulations are usually performed, because the parameters required to handle the
experimental data (detector efficiency, energy spectrum of the incoming neutrons, etc.)
are usually unknown or difficult to determine experimentally.
Table~\ref{flux_measurement} summarizes the results of the neutron flux
measurements at the Gran Sasso laboratory. One measurement was performed in hall C by the
ICARUS collaboration \cite{arneodo}, while all others were performed in hall A of the
laboratory. Most of the measurement results are given as
the integral flux in a large energy bin. Although different energy binning was used in
these measurements, one can see that the results vary rather widely. Only two measurements
\cite{arneodo,belli} give some information on the spectral
shape. However, for some purposes a more detailed spectrum is
needed.

The total flux of \cite{arneodo} is basically consistent with the
result of \cite{belli} for $E > 1$ MeV, while the energy spectrum
as a whole is significantly harder. The results from
\cite{arneodo} show important distortions with respect to the
spectrum that was previously generally assumed in literature,
i.e. a spectrum produced mainly by spontaneous fission.
The difference is seen at high energies. To search for a possible
explanation the authors of \cite{arneodo} considered the
production of neutrons via ($\alpha$,n) reactions inside the rock
and computed the `thin target' spectrum of neutrons produced by
the reaction $^{17}$O($\alpha$,n)$^{20}$Ne only. The ($\alpha$,n)
and spontaneous fission spectra were then used as input for the
MCNP (Monte Carlo N-Particles) code to obtain the effective
neutron flux in hall C. They found that their measurements and
simulations are in very good agreement and that ($\alpha$,n)
reactions are the main source of the high energy neutron flux in
the laboratory. However, according to \cite{gerbier} the
uncertainties quoted in \cite{arneodo} have been underestimated.
They should be multiplied by $\sqrt{3}$ to get the
proper uncertainties, due to the fact that the background run in
this experiment is about three times shorter than the data run.
Then all data points shown in \cite{arneodo} would be
insignificant and the agreement between the thin target calculation
and the measurement found here would be merely accidental.

Based on Monte Carlo simulations, this work is devoted to
reinvestigate the neutron flux from ($\alpha$,n) and fission
reactions in the rock and concrete of the Gran Sasso laboratory.
The results offer an independent check to the present experimental
data and give detailed information on the shape of the spectrum.

\section{Calculation of Neutron Production by ($\alpha$,n)
and Fission Reactions in the Rock and Concrete}
\subsection{Composition and Activity of the Rock and Concrete}
Gran Sasso rock consists mainly of CaCO$_3$ and MgCO$_3$, with a
density of 2.71$\pm0.05$ g/cm$^{3}$ \cite{catalano}. The weight
percentage of the elements is given in Table~\ref{rock}. Due to
the presence of a certain type of rock, called ``roccia marnosa
nera", the contaminations of $^{238}$U and $^{232}$Th in hall A
rock are about ten and thirty times higher respectively than those
of hall C \cite{bellotti} as shown in
Table~\protect\ref{activity1}.

Since there are no data on the chemical composition of Gran Sasso
concrete available in literature, several samples were taken from
different places in the laboratory. The concrete samples were then
analyzed at the laboratory of the \textit{Lehrstuhl f\"{u}r
Baustoffkunde und Werkstoffpr\"{u}fung} at the department of civil
engineering of the Technische Universit\"{a}t M\"{u}nchen. No
significant variations were found in the chemical composition of
the samples, which leads to the conclusion that all halls in Gran
Sasso are layered with the same type of concrete. The typical
water content in the concrete is 12\%, with a possible variation
of 4\% at most (in most cases the variation is smaller). The
weight percentage of elements in concrete with 8\% water content
(hereafter ``dry concrete") is shown in Table~\ref{concrete}. The
$^{238}$U and $^{232}$Th contaminations are $1.05\pm0.12$ ppm and
$0.656\pm0.028$ ppm \cite{bellini} respectively. The density is
between 2.3 and 2.5 g/cm$^{3}$, depending on the assumed water content.

\subsection{Neutron Production by Spontaneous Fission}
There are mainly three nuclides in nature that undergo spontaneous
fission: $^{238}$U, $^{235}$U and $^{232}$Th. Only neutrons
produced by spontaneous fission of $^{238}$U are considered here,
because of the long fission half life of the other two nuclides
compared to that of $^{238}$U. The  spectrum of the emitted
neutrons follows the Watt spectrum:
\begin{equation}
N(E) = C \mbox{exp}\,(-E/a)\mbox{sinh}\,(bE)^{1/2} \label{eq:watt}
\end{equation}

\begin{figure}[p]
\begin{center}
\includegraphics*[width=3.2in]{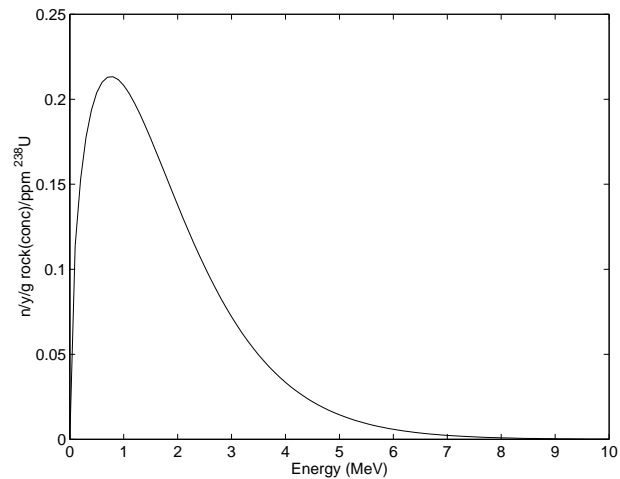}
\end{center}
\caption{Energy spectrum of neutrons from spontaneous fission of
$^{238}$U.} \label{fission}
\end{figure}

In this work the Watt spectrum parameters of the Los Alamos model
results \cite{madland} have been used, where $a=0.7124\,$MeV and
$b=5.6405\,$MeV$^{-1}$. The rate of spontaneous fission of
$^{238}$U is 0.218/year/g of rock (concrete) for 1\,ppm of
$^{238}$U and the average number of neutrons emitted per fission
event is 2.4$\pm$0.2 \cite{littler}. This gives 0.52
neutrons/year/g of rock (concrete)/ppm $^{238}$U. This number
multiplied by the $^{238}$U activities in the rock/concrete would
give 3.54, 0.22 and 0.34 neutrons/year/g in the rock of hall A, B
and C respectively and 0.55 neutrons/year/g in the concrete.
Figure~\protect\ref{fission} shows the neutron energy distribution
due to spontaneous fission of $^{238}$U.

\subsection{Neutron Production by ($\alpha$,n) Reactions}
Uranium, thorium, and their daughter products decay by emitting
$\alpha$ and $\beta$ particles. In the rock (concrete)
$\alpha$-particles can interact especially with light elements and
produce neutrons through ($\alpha$,n) reactions.

The yield of neutrons per $\alpha$-particle for an individual
element depends on the ($\alpha$,n) interaction cross section
(which is energy dependent), and on the energy loss of
$\alpha$-particles in a medium made of that element. In this work
the thick target yield of ($\alpha$,n) reactions was used instead
of the thin target yield used in \cite{arneodo}. The thick target
yield of the ($\alpha$,n) reaction for an individual element
\textit{j} in which the $\alpha$-particle has a range $R$ can be
written as \cite{feige}:
\begin{equation}
Y_{j}=\int_{0}^{R}n_{j}\,\sigma_{j}(E)\,dx
\label{Eq:thick_yield}
\end{equation}
where $n_{j}$ is the number of atoms per unit volume of element
$j$, and $\sigma_{j}$ is the microscopic ($\alpha$,n) reaction
cross section for an $\alpha$-particle energy $E$. Transforming
the right side of Eq.~\protect\ref{Eq:thick_yield} into an
integral over energy gives:
\begin{eqnarray}
Y_{j}&=&\int_{0}^{E_{i}}\frac{n_{j}\,\sigma_{j}(E)}{-(dE/dx)}dE\nonumber\\
&=&\int_{0}^{E_{i}}\frac{n_{j}\,\sigma_{j}(E)}{\rho_{j}S_{j}^{m}(E)}dE\nonumber\\
&=&\int_{0}^{E_{i}}\frac{n_{j}\,\sigma_{j}(E)N_{A}}{n_{j}A_{j}S_{j}^{m}(E)}dE\nonumber\\
&=&\frac{N_{A}}{A_{j}}\int^{E_{i}}_{0}\frac{\sigma_{j}(E)}{S^{m}_{j}(E)}\,dE
\label{eq:yield_isolation}
\end{eqnarray}
where $E_{i}$ is the initial $\alpha$-energy, $N_{A}$ is Avogadro's number,
$A_{j}$ is the atomic mass and $\sigma_{j}$ and
$S^{m}_{j}$ are the ($\alpha$,n) cross section and the mass
stopping power respectively, which are energy dependent.

Neutron yields from individual elements can be used to calculate
the total yield in a chemical compound or mixture
\cite{feige,west,bulanenko}. The following assumptions are usually
made in such a calculation:
\begin{list}
~\protect\item[(i)]  the compound is a homogeneous mixture of its
constituent elements
~\protect\item[(ii)] Bragg's law of
additivity for stopping power holds for the compound
~\protect\item[(iii)] the ratio of an element's stopping power to
the total stopping power of the compound is independent of the
$\alpha$-particle energy.
\end{list}

The validity of (iii) has been discussed by several authors.
Although for each element the mass stopping power $S_{j}^{m}$
decreases with energy, Feige \cite{feige} found that the mass
stopping power ratio for any pair of elements does not change by
more than $\pm\,$4\% between 5.3 MeV and 8.8 MeV. Heaton
\textit{et al.} \cite{heaton} showed that above 3 MeV this
approximation introduces an uncertainty of less than 5\% in the
neutron yield.

Under those assumptions the neutron yield of element \textit{j} in
the compound or mixture with initial $\alpha$-particle energy
$E_{i}$ can be written as:
\begin{eqnarray}
Y_{i,j,mix}&=&\int_{0}^{E_{i}}\frac{n_{j}\,\sigma_{j}}{\sum_{j}\rho_{j}S_{j}^{m}}dE\nonumber\\
&=&\int_{0}^{E_{i}}\frac{\rho_{j}S_{j}^{m}}{\sum_{j}\rho_{j}S_{j}^{m}}\frac{n_{j}\,
\sigma_{j}(E)}{\rho_{j}S_{j}^{m}}dE\nonumber\\
&=&\frac{\rho_{j}S_{j}^{m}(E_{0})}{\sum_{j}\rho_{j}S_{j}^{m}(E_{0})}\int_{0}^{E_{i}}\frac{n_{j}\,
\sigma_{j}(E)}{\rho_{j}S_{j}^{m}}dE\nonumber\\
&=&\frac{M_{j}S_{j}(E_{0})}{\Sigma_{j}M_{j}S_{j}(E_{0})}Y_{j}(E_{i})
\end{eqnarray}
where $M_{j}$ is the mass fraction of element \textit{j} in the
mixture, $E_{0}$ is a chosen reference energy (8 MeV in this
work), $S_{j}^{m}$ is the mass stoping power and $Y_{j}(E_{i})$ is
the neutron yield of element \textit{j} in isolation (see
Eq.~(\ref{eq:yield_isolation})). Thus the ($\alpha$,n) yield of a
compound (mixture) is the sum of the yields of its elements
weighted by the relative contributions of the elements to the total
stopping power of the compound. The use of 8 MeV mass stopping
power was selected because the overwhelming contribution to the
neutron yield comes from high energy $\alpha$-particles and the
relative stopping power of elements are nearly independent of the
$\alpha$-energy in this energy region \cite{heaton2}. The mass
stopping powers at 8 MeV of elements used in this work are as
those used in \cite{heaton}. In this work the data on neutron
yield of elements in isolation, in units of neutron/$\alpha$, is
taken from a compilation by Heaton et al. \cite{heaton2,heaton3}.

Each $\alpha$ emitter in the $^{238}$U and $^{232}$Th decay chains
emits $\alpha$'s at a certain energy, which was used as the
initial energy in the neutron yield calculation. The neutron yield
of each element with certain initial energy was then multiplied by
the branching ratio, the number of $\alpha$'s emitted by each
emitter per unit time, and the concentration of $^{238}$U and
$^{232}$Th in the rock (concrete). It is assumed in this work,
that $^{238}$U and $^{232}$Th are in secular equilibrium with
their daughter products. The total neutron production rate for
each element was calculated by summing up the contribution of all
alpha emitters.

In Table~\ref{hallyield} and Table~\ref{concyield} neutrons
produced per unit mass per year  by ($\alpha$,n) reactions with
the elements of the rock in hall A, hall C and also in dry
concrete are presented. It is seen that fission (see discussion in
the previous section) and ($\alpha$,n) reactions contribute more
or less equally to the total production rates both in the rock and
in the concrete. Half of the total ($\alpha$,n) neutron production
in the rock comes from interactions of $\alpha$ particles with
magnesium, which comprises only less than 6\% of the weight
percentage of the rock, whereas oxygen with almost 50\% weight
percentage contributes to only about 20\% of the production rate.
Due to the higher activity of the hall A rock the ($\alpha$,n)
neutron production in the rock of this hall is more than ten times
higher than in the hall C rock. In the concrete Na, Al and Mg
contribute significantly in spite of their minor weight
percentages. The production rate per unit mass in the wet concrete
(16\% water content) is slightly smaller than in the dry concrete,
but this is merely due to the difference in the densities. The
volume of concrete remains, while the mass changes with the water
content. Given per unit volume the production rates in dry and wet
concrete are the same.

The energy of the emitted neutron is dependent on the $\alpha$
energy, the reaction energy $Q$, and the neutron emission angle.
It was calculated under the following assumptions:
\begin{list}
~\protect\item[(i)] the interaction take place at the initial
$\alpha$ energy
~\protect\item[(ii)] the neutron is emitted at
$90^{0}$
~\protect\item[(iii)] the residual nucleus is produced in
its ground state.
\end{list}
Under these assumptions, the neutron energy can be determined by using
the simple Eq.~(\ref{eq:energy}):
\begin{equation}
E_n = \frac{MQ+E_\alpha(M-M_\alpha)}{(M_n+M)} \label{eq:energy}
\end{equation}
Here $M$ is the mass of the final nucleus, $M_n$ and $M_\alpha$
are the masses of neutron and the $\alpha$ particle respectively,
and $E_\alpha$ is the initial $\alpha$ energy.

The threshold energy $E_{th}$ for an ($\alpha$,n) reaction is the
minimum kinetic energy the impinging $\alpha$ particle must have
(in the laboratory system) in order to make the reaction
energetically possible. For endothermic reactions, the threshold
energy is:
\begin{equation}
E_{th}=-[(M_{n}+M_{\alpha})/M_{1}]Q
\end{equation}
where $M_{1}$ and $M_{\alpha}$ are the masses of the target
nucleus and of $\alpha$ particle respectively. The threshold
energy is zero for exothermic reactions, i.e. reactions with
positive $Q$.

\begin{figure}[p]
\begin{center}
\includegraphics*[width=3.3in]{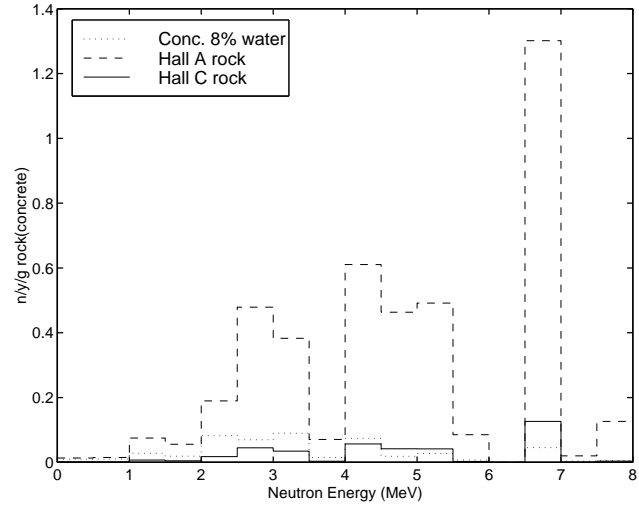}
\end{center}
\caption{Energy spectra of emitted neutrons from $(\alpha$,n) reactions as
used in the simulations.} \label{plotconc}
\end{figure}

The highest energy among naturally emitted relevant $\alpha$
particles is 8.79 MeV, which comes from the decay of $^{212}$Po.
Hence, for some elements in the rock/con\-crete there are isotopes
that can not participate in the ($\alpha$,n) reactions because of
their high $E_{th}$ (e.g. $^{16}$O, $^{28}$Si, $^{24}$Mg, and
$^{40}$Ca). For each element calculations of neutron energies were
therefore done for all isotopes that are not closed for
$(\alpha$,n) interactions. Then, neutron mean energies of elements
were calculated according to the relative abundances of the
``open" isotopes for different $\alpha$ energies. Finally, the
yields of all elements were summed up in 0.5\,MeV energy bins to
get the energy spectra of neutrons from $(\alpha$,n) reactions as
shown in Figure~\protect\ref{plotconc} for hall A rock, hall C
rock, and dry concrete. While neutrons below 4\,MeV are mainly
produced by spontaneous fission, ($\alpha$,n) reaction is the main
contributor in the production of neutrons with higher energy.
Neutrons with energy above 6\,MeV are contributions of
magnesium and carbon.

\section{Flux of ($\alpha$,n) and Fission Neutrons at LNGS}
To get the flux of low energy neutrons inside the halls, the Monte
Carlo code MCNP4B (Monte Carlo N-Particles version 4B) from Los
Alamos \cite{briesmeister} was used to transport neutrons produced
by fission and $(\alpha$,n) reactions as calculated above through the rock and
concrete and scatter them inside the halls.

Table~\protect\ref{typ_depth} summarizes the typical depths of
neutrons (the depths where $\sim$ 63\% of neutrons come from, the
$1/e$ length) produced in hall A rock and in dry concrete, and
entering hall A (before scattering inside the hall) with any
energy and in addition with energy $E>1$ MeV. Neutrons with $E>1$
MeV come mainly from the first 7\,cm and 13\,cm of concrete and
rock respectively. As the thickness of the concrete layer at the
Gran Sasso laboratory is not less than 30\,cm (in some places even
around 1\,m), the bulk of the total flux at the laboratory is
given by neutrons produced in the concrete.

In the further simulations performed to get the neutron flux
inside the halls, the thickness of the concrete layer in the
laboratory has been set to 45\,cm below the floor and 35\,cm
elsewhere, and 2\,m of rock have been taken into account. The
calculated fluxes of neutrons (after scattering inside the halls)
are shown in Table~\protect\ref{nflux_calcu} for hall A with 8\%
(dry) and 16\% (wet) water content in the concrete and  for hall C
with dry concrete, together with the measurement in Hall A by
Belli et al. \cite{belli} as a comparison. The measured flux
between $1-\,2.5\,$MeV is taken from \cite{arneodo}. We calculated
the errors in this energy bin by assuming that the relative errors
in 1\,keV$\,-\,$1\,MeV and $1-2.5\,$MeV bins are equal.

Uncertainties in the neutron flux in our calculations can be
attributed to the uncertainties in the neutron production rate
from ($\alpha$,n)/fission reactions and
the energies of the emitted neutrons.\\
The major sources of uncertainty in the total ($\alpha$,n) neutron
yield of the rock and concrete are the uncertainties in the
stopping powers, elemental neutron yield per $\alpha$ particle and
material compositions. We adopt the 5\% uncertainty in the
stopping power data as stated in \cite{heaton2}. The assumption
that the ratio of an element's stopping power to the total
stopping power of the compound is independent of the
$\alpha$-particle energy introduces another uncertainty of about
5\% in the neutron yield (see Subsection 2.3). The deviation from
Bragg's Rule, which was observed in compounds in gaseous phase and
hydrocarbons, does not produce significant contribution to the
uncertainty of the neutron yield calculation (see \cite{heaton2}
and references therein). An additional 5\% uncertainty coming from
the Heaton's compilation of elemental neutron yield per $\alpha$
above 3.7 MeV \cite{heaton2} is taken into account. For the
determination of the material composition we assume an uncertainty
of 1-2\%, as is common using modern analytical instruments.
Combining these uncertainties in quadrature, the uncertainty of
the ($\alpha$,n) yield from a compound is roughly 9\%. The major
uncertainty in the fission neutron yield of  about 8\% is due to
uncertainty in the average number of neutrons emitted per fission
\cite{littler}. The total number of neutrons produced by fission
and ($\alpha$,n) in the rock/concrete at the Gran Sasso laboratory
depends eventually on the $^{238}$U and $^{232}$Th contamination.
Taking into account the uncertainties in the activities and the
aforementioned uncertainties in the neutron yield would give an
over all uncertainties of about 15\% and 20\% in the total neutron
fluxes in hall A and hall C respectively, due to uncertainties in
the neutron production from ($\alpha$,n)/fission reactions.

Jacobs and Liskien \cite{jacobs} reported energy spectra of
neutrons produced by $\alpha$-particles with energies of 4, 4.5, 5
and 5.5 MeV in thick targets of light elements (all elements
important for us are available here, except Na). The difference
between our calculation of the energies of emitted neutrons and
the average energies in \cite{jacobs} are generally less than 0.5
MeV. Only for Mg a difference of 1.5 MeV is found. To estimate the
uncertainty in the neutron flux due to our assumptions in
determining the energies of the emitted neutrons (see Subsection
2.3 and Figure~\protect\ref{plotconc}) we have repeated the
simulations of neutron propagation, using the average neutron
energies reported in \cite{jacobs} where available. For
$\alpha$-energies above 5.5 MeV we reduced our calculated emitted
neutron energies by 0.5 MeV (1.5 MeV for Mg). For elements which
are not available from \cite{jacobs} we reduced the energies of
emitted neutrons by 0.5 MeV for all $\alpha$-energies. For the
spontaneous fission of $^{238}$U no modification is needed since
the spectrum is well understood. We find that the total neutron
fluxes at the hall A and C change by about 10\% due to these
alterations, and also the spectral shape changes. Combining this
uncertainty with the aforementioned uncertainties in the neutron
production, we estimate the over all systematic uncertainty of the total flux to be
around 20\% (slightly higher for hall C). Energy dependent
uncertainties are reported in Table~\protect\ref{nflux_calcu}.

The total flux and the integral flux above 1\,MeV in hall A are
consistent with those of the measurement \cite{belli} if the
concrete in hall A is dry. The simulated flux differs from the
measured one in the chosen bins below 1\,MeV. In our simulations
we have assumed that there is nothing inside the hall and neutrons
can only be scattered by the walls. In reality as in the
measurement, neutrons coming from the rock/concrete can be
scattered by anything inside the hall before they eventually come
into the experimental setup. Those neutrons are moderated, raising
the flux in the lowest energy bin.

Within the estimated uncertainties the total flux in hall C is only
slightly less than in hall A for the case of dry concrete, although the neutron production
rate in hall C rock is more than ten times lower than that of hall
A; above 1\,MeV the fluxes in the two halls are in agreement. This is due to
the concrete, which indeed reduces the neutron flux from the rock
significantly so that neutrons coming into the halls are mainly
those produced in the concrete layer.

Table~\protect\ref{nflux_calcu} shows that the neutron flux
depends on the humidity of the environment. The flux in Hall A is
lower if the concrete is wet than if it is dry (8\% and 16\% water
content respectively). As mentioned in the previous section, the
8\% difference in the water content of concrete does not lead to
different neutron production rates. The effect seen in the flux
here is caused only by moderation. Wet concrete moderates neutrons
more effectively than dry concrete due its higher hydrogen
content. The fluxes obtained for dry and wet concrete here show
the maximum possible variation for the water content of concrete.
A more realistic variation of the water content of ($12\pm1$)\% results
consequently in smaller flux variation. To quantify this effect
and to see whether it is a seasonal phenomenon, it is necessary to
monitor the water content of the concrete for at least one year.

\begin{figure}[p]
\begin{center}
\includegraphics*[width=3.6in]{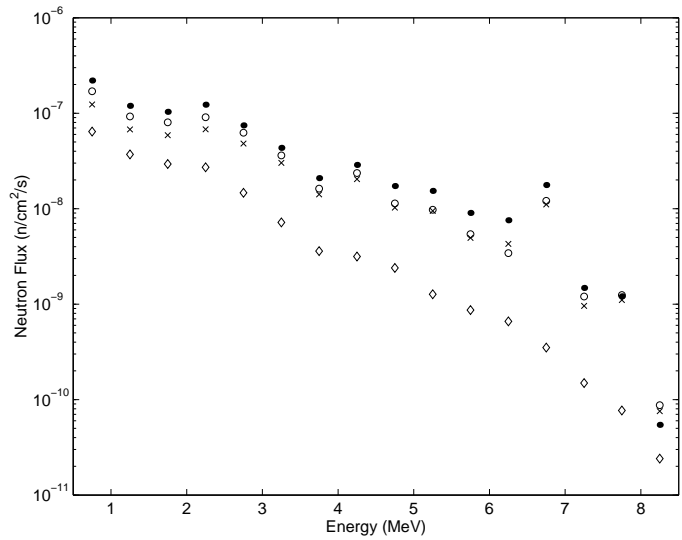}
\end{center}
\caption{Neutron flux at the Gran Sasso laboratory,
$\mbox{\large{$\bullet$}}$: hall A, dry concrete,
$\mbox{{$\times$}}$: hall A, wet concrete, $\lozenge$: hall A, dry concrete,
fission reactions only and $\mbox{\large{$\circ$}}$: hall C, dry
concrete. Each point shows the integral flux in a 0.5 MeV energy
bin.} \label{compare_spectrum}
\end{figure}

Detailed spectra of neutrons in hall A and hall C are shown in
Figure~\protect\ref{compare_spectrum} for neutron energy above
0.5\,MeV. Each point shows the integral flux in a 0.5 MeV energy
bin. The contribution of ($\alpha$,n) makes the spectra in both
halls differ from the spectrum expected for neutrons produced by
fission reactions only, as was previously generally assumed,
especially at high energies.

\section{Conclusion}
We have discussed the flux of neutrons induced by radioactivity in
the rock and concrete surrounding the Gran Sasso laboratory. The
flux is dominated by neutrons produced in the concrete layer and
therefore does not vary much from hall to hall. It can be expected
that as well for other underground laboratories the neutron flux
originates mainly from the concrete and not from the rock
material. A more detailed spectrum compared to that from
measurements has been obtained. The spectrum differs from the
spectrum expected for neutrons produced by fission reactions only,
especially at high energies, due to the contribution of
($\alpha$,n) neutrons. We also have shown the dependence of the
neutron flux on the humidity of the concrete. Our results for the
case of hall A with dry concrete are in good agreement with the
experimental data from \cite{belli}.

\section{Acknowledgement}
H. Wulandari thanks the \textit{Deutscher Akademischer Austausch Dienst} (DAAD)
for the financial support of her PhD work.


\clearpage


\begin{table}[p]
\caption{Neutron flux measurements at the Gran Sasso laboratory
reported by different authors. In analyzing their experimental
data with Monte Carlo simulations, Belli et al.\,~\protect\cite{belli} have used two different hypothetical
spectra: flat, and flat plus a Watt fission spectrum. This leads
to the upper and lower data sets shown for
ref.~\protect\cite{belli} respectively.} {\tabcolsep1pt
\begin{tabular}{|c|c|c|c|c|c|c|}
\hline
 E interval &\multicolumn{6}{c|}{Neutron Flux (10$^{-6}$cm$^{-2}$s$^{-1}$)}\\
\cline{2-7}
(MeV) & Ref.\,~\protect\cite{aleksan}& Ref.\,~\protect\cite{arneodo}& Ref.\,~\protect\cite{belli}&
Ref.\,~\protect\cite{bellotti}& Ref.\,~\protect\cite{cribier}& Ref.\,~\protect\cite{rindi}\\
\hline \hline
10$^{-3}-0.5$ & & & & & & \\
\cline{1-1} \cline{7-7}
$0.5-1$ &  & & 0.54$\pm$0.01& & & \\
 \cline{1-1} \cline{3-3}
$1-2.5$ & & 0.14$\pm$0.12 & (0.53$\pm$0.08)& & &\\
\cline{1-1} \cline{3-5} \cline{6-6}
$2.5-3$ & & 0.13$\pm$0.04 &0.27$\pm$0.14 & & &\\
\cline{1-2}
$3-5$ & & & (0.18$\pm$0.04)& & & 2.56$\pm$0.27\\
\cline{1-1} \cline{3-4}
$5-10$ & & 0.15$\pm$0.04 &0.05$\pm$0.01 & & & \\
 & & &(0.04$\pm$0.01) & 3.0$\pm$0.8& 0.09$\pm$0.06& \\
\cline{1-1} \cline{3-4} \cline{7-7}
$10-15$ & 0.78$\pm$0.3 & $(0.4\pm0.4)\!\cdot\!10^{-3}$ & $(0.6\pm0.2)\!\cdot\!10^{-3}$& & & \\
 & & &($(0.7\pm0.2)\!\cdot\!10^{-3}$) & & & \\
\cline{1-1} \cline{3-4}
$15-25$ &  & & $(0.5\pm0.3)\!\cdot\!10^{-6}$& & & \\
 & & &($(0.1\pm0.3)\!\cdot\!10^{-6}$) & & & \\
\hline
\end{tabular}
\label{flux_measurement}}
\end{table}

\begin{table}[p]
\caption{Chemical composition of LNGS rock.} {\tabcolsep6.0pt
\begin{tabular}{|c|c|c|c|c|c|c|c|}
\hline
Element& C& O& Mg& Al & Si&  K & Ca\\
\hline
\% Weight& 11.88& 47.91& 5.58& 1.03& 1.27&  1.03& 30.29\\
\hline
\end{tabular}
\label{rock}}
\end{table}

\begin{table}[p]
\caption{$^{238}$U and $^{232}$Th activities in LNGS rock.}
{\tabcolsep8.0pt
\begin{tabular}{|c|c|c|}
\hline
Hall& \multicolumn{2}{|c|}{Activities (ppm)}\\
\cline{2-3}
 & $^{238}$U& $^{232}$Th\\
\hline\hline
A& $6.80\pm0.67$& $2.167\pm0.074$\\
B& $0.42\pm0.10$& $0.062\pm0.020$\\
C& $0.66\pm0.14$& $0.066\pm0.025$\\
\hline
\end{tabular}
\label{activity1}}
\end{table}


\begin{table}[p]
\caption{Chemical composition of LNGS dry concrete.}
{\tabcolsep3.5pt
\begin{tabular}{|c|c|c|c|c|c|c|c|c|c|c|c|c|c|}
\hline
Element &H & C& O & Na &Mg& Al & Si& P& S & K & Ca& Ti& Fe   \\
\hline \% Weight &0.89& 7.99& 48.43 &0.6& 0.85 &0.9 & 3.86& 0.04&
0.16 & 0.54& 34.06 &0.04 &0.43   \\
\hline
\end{tabular}
\label{concrete}}
\end{table}

\begin{table}[p]
\begin{minipage}{2.4 in}
\caption{Neutron yields from $(\alpha,n)$  interactions in the
rock.} {\tabcolsep4.0pt
\begin{tabular}{|c|c|c|}
\hline
Element& \multicolumn{2}{|c|}{Total elemental yield} \\
 & \multicolumn{2}{|c|}{(n/y/g\,rock)} \\
 \cline{2-3}
 & Hall A & Hall C\\
\hline \hline
C & 4.60E-1& 4.07E-2\\
O & 8.8E-1 &7.90E-2\\
Mg & 2.31E+0& 2.04E-1\\
Al & 3.50E-1& 3.05E-2\\
Si & 6.00E-2 &5.21E-3\\
K & 9.00E-2& 7.60E-3\\
Ca & 2.40E-1& 2.05E-2\\
\hline
Total yield & 4.38E+0 & 3.88E-1\\
\hline
\end{tabular}
\label{hallyield}}
\end{minipage}\hspace*{25pt}
\begin{minipage}{2.7 in}
\caption{Neutron yields from $(\alpha$,n) interactions in dry
concrete (8\% water content).} {\tabcolsep10pt
\begin{tabular}{|c|c|}
\hline
Element& Total elemental yield \\
 & (n/y/g\,concrete)\\
\hline \hline
C & 5.24E-2 \\
O & 1.50E-1\\
Na &9.65E-2 \\
Mg & 6.07E-2\\
Al & 5.35E-2\\
Si & 3.16E-2\\
P &4.35E-4 \\
S & 3.36E-4 \\
K & 8.31E-3 \\
Ca & 4.95E-2 \\
Ti & 3.35E-4\\
Fe & 9.53E-4 \\
\hline
Total yield & 5.05E-1\\
\hline
\end{tabular}
\label{concyield}}
\end{minipage}
\end{table}

\begin{table}[p]
\caption{Typical depths (in cm) of ($\alpha$,n) and fission
neutrons produced in hall A rock and dry concrete. ``Energy"
represents the energy of neutrons entering hall A before
scattering in the hall.} {\tabcolsep15.0pt
\begin{tabular}{|c|c|c|c|c|}
\hline
Energy& \multicolumn{2}{c}{Hall A rock} &\multicolumn{2}{|c|}{Dry concrete}\\
\cline{2-3} \cline{4-5}
 & Fission& ($\alpha$,n)&  Fission& ($\alpha$,n)\\
\hline\hline
all energies& 23 & 24 & 10.5 &12\\
\hline
$> 1$ MeV& 10& 13 &6.5 &6.5\\
\hline
\end{tabular}
\label{typ_depth}}
\end{table}

\clearpage

\begin{table}[p]
\caption{Simulated neutron flux in hall A with dry (8\% water
content) and wet (16\% water content) concrete, in hall C with dry
concrete and the measurements by Belli et al.~\protect\cite{belli}
for the two hypothetical spectra mentioned in the caption of
Table~\ref{flux_measurement}.} {\tabcolsep1.pt
\begin{tabular}{|c|c|c|c|c|c|}
\hline
Energy & \multicolumn{5}{|c|}{Neutron Flux ($10^{-6}\mbox{n/cm}^{2}\mbox{s}$)}\\
\cline{2-6}
(MeV)& \multicolumn{2}{|c|}{Measurement in hall A \cite{belli}}& \multicolumn{3}{|c|}{MC Simulations, this work} \\
\cline{2-6}
& Flat & Flat+Watt Spect.& Hall A,dry &Hall A,wet &Hall C,dry \\
\hline\hline
$0-5.10^{-8}$ & 1.08$\pm$0.02& 1.07$\pm$0.05& 0.53$\pm$0.36 & 0.24$\pm$0.15& 0.24$\pm$0.17\\[-1ex]
\hline
$5.10^{-8}-10^{-3}$ & 1.84$\pm$0.20& 1.99$\pm$0.05& 1.77$\pm$0.45 & 0.71$\pm$0.19 &0.93$\pm$0.33\\[-1ex]
\hline
$10^{-3}-2.5$ & 0.54$\pm$0.01& 0.53$\pm$0.08& 1.22$\pm$0.32 & 0.57$\pm$0.16&0.91$\pm$0.32 \\[-1.5ex]
($1\,-\,2.5$) & (0.38$\pm$0.01)$\!$\footnotemark&(0.38$\pm$0.06)$^{1,}\!$\footnotemark &(0.35$\pm$0.12)& (0.18$\pm$0.06) & (0.27$\pm$0.12) \\[-1ex]
\hline
$2.5\,-\,5$ & 0.27$\pm$0.14& 0.18$\pm$0.04&0.18$\pm$0.05& 0.12$\pm$0.04 & 0.15$\pm$0.05 \\[-1ex]
\hline
$5\,-\,10$ & 0.05$\pm$0.01& 0.04$\pm$0.01&0.05$\pm$0.02& 0.03$\pm$0.02 & 0.03$\pm$0.01 \\[-1ex]
\hline\hline
Total Flux &3.78$\pm$0.25 &3.81$\pm$0.11 &3.75$\pm$0.67 &1.67$\pm$0.29 &2.26$\pm$0.49 \\[-1.5ex]
Flux(E$>1\,$MeV) &0.70$\pm$0.14 &0.60$\pm$0.07 &0.58$\pm$0.13 &0.33$\pm$0.07 &0.45$\pm$0.13 \\
\hline
\end{tabular}
\addtocounter{footnote}{-2}
\myFootnote{Taken from~\protect\cite{arneodo}.}\,\,
\myFootnote{Uncertainties are assumed to be equal below and above
1\,MeV.} \label{nflux_calcu}}
\end{table}

\pagebreak[4]

\end{document}